\documentclass{aa}
\usepackage{graphicx}
\usepackage{txfonts}
\def\swift{{\em Swift\/ }}
\def\xmm{{\em XMM-Newton\/ }}

\def\ergcms{\mbox{ erg cm$^{-2}$ s$^{-1}$}}

\newcommand{\arcm}{\hbox{$^\prime$}}
\newcommand{\arcs}{\hbox{$^{\prime\prime}$}}

\newcommand{\ls}
{\mathrel{\hbox{\rlap{\hbox{\lower4pt\hbox{$\sim$}}}\hbox{$<$}}}}
\newcommand{\gs}
{\mathrel{\hbox{\rlap{\hbox{\lower4pt\hbox{$\sim$}}}\hbox{$>$}}}}

\begin{document}

   \title{The exceptionally extended flaring activity in the X--ray afterglow of GRB 050730
          observed with Swift and XMM-Newton
         }


   \author{
           M.~Perri\inst{1,2},
           D.~Guetta\inst{2},
           L.A.~Antonelli\inst{1,2},
           A.~Cucchiara\inst{3},
           V.~Mangano\inst{4},
           J.~Reeves\inst{5}, 
           L.~Angelini\inst{5}, 
           A.P.~Beardmore\inst{6},
           P.~Boyd\inst{5}, 
           D.N.~Burrows\inst{3}, 
           S.~Campana\inst{7}, 
           M.~Capalbi\inst{1,2}, 
           G.~Chincarini\inst{7,8},
           G.~Cusumano\inst{4}, 
           P.~Giommi\inst{1,9},
           J.E.~Hill\inst{5}, \\
           S.T.~Holland\inst{5,10},
           V.~La Parola\inst{4}, 
           T.~Mineo\inst{4}, 
           A.~Moretti\inst{7}, 
           J.A.~Nousek\inst{3}, 
           J.P.~Osborne\inst{6}, 
	   C.~Pagani\inst{3}, 
           P.~Romano\inst{7}, \\
           P.W.A.~Roming\inst{3},
           R.L.C.~Starling\inst{6},
           G.~Tagliaferri\inst{7}, 
           E.~Troja\inst{4,6}, 
           L.~Vetere\inst{1,3} and
           N.~Gehrels\inst{5}
          }
          
   \offprints{\email{perri@asdc.asi.it}}

   \institute{ASI Science Data Center, Via Galileo Galilei, 
              I-00044 Frascati, Italy
       \and INAF -- Astronomical Observatory of Rome, 
              Via Frascati 33, I-00040 Monte Porzio Catone (Rome), Italy
       \and Department of Astronomy \& Astrophysics, Pennsylvania State 
              University, University Park, PA 16802, USA
       \and INAF -- Istituto di Astrofisica Spaziale e Fisica Cosmica,
              Sezione di Palermo, Via La Malfa 153, I-90146 Palermo, Italy
       \and NASA/Goddard Space Flight Center, Greenbelt, MD 20771, USA
       \and Department of Physics \& Astronomy, University of Leicester,
              Leicester LE1 7RH, UK
       \and INAF -- Astronomical Observatory of Brera, 
              Via Bianchi 46, I-23807 Merate, Italy
        \and Universit\`a degli Studi di Milano-Bicocca, Dipartimento 
              di Fisica, Piazza delle Scienze 3, I-20126 Milano, Italy
       \and Agenzia Spaziale Italiana, Unit\`a Osservazione dell'Universo,
              Viale Liegi 26, I-00198 Roma, Italy
        \and Universities Space Research Association, 10211 Wincopin Circle, Suite 500, 
              Columbia, MD, 21044-3432, USA
             }

   \authorrunning{M. Perri et al.}
   \titlerunning{The exceptionally extended flaring activity in the X--ray afterglow of GRB 050730}
   \date{Received: 11 August 2006 / Accepted: 27 March 2007}

\abstract
{}
{
We observed the high redshift ($z=3.969$) GRB 050730 with \swift and \xmm to study its prompt and afterglow emission.
}
{
We carried out a detailed spectral and temporal analysis of \swift and \xmm observations.
}
{
The X--ray afterglow of GRB 050730 was found to decline with time with superimposed intense flaring activity 
that extended over more than two orders of magnitude in time. Seven distinct re-brightening events starting from 236 s 
up to 41.2 ks after the burst were observed. 
The underlying decay of the afterglow was well described by a double broken power-law model with breaks at 
$t_1 = 237\pm 20~\mathrm{s}$ and $t_2 = 10.1_{-2.2}^{+4.6}~\mathrm{ks}$. The temporal decay slopes before, between 
and after these breaks were $\alpha_1 = 2.1\pm 0.3$, $\alpha_2 = 0.44_{-0.08}^{+0.14}$ and 
$\alpha_3 = 2.40_{-0.07}^{+0.09}$, respectively. 
The spectrum of the X--ray afterglow was well described by a photoelectrically absorbed power-law with 
an absorbing column density $N^{z}_\mathrm{H}$=(1.28$^{+0.26}_{-0.25})\times10^{22}$\,cm$^{-2}$ 
in the host galaxy. Evidence of flaring activity in the early UVOT optical afterglow, simultaneous with that 
observed in the X--ray band, was found. 
Strong X--ray spectral evolution during the flaring activity was present. The rise and decay power-law slopes 
of the first three flares were in the range 0.8--1.8 using as zero times the beginning and the peak of the 
flares, respectively.
In the majority of the flares (6/7) the ratio $\Delta t/t_{\rm p}$ between the duration of the event and the time 
when the flare peaks was nearly constant and $\sim$\,0.6--0.7. 
We showed that the observed spectral and temporal properties of the first three flares are consistent with 
being due both to high-latitude emission, as expected if the flares were produced by late internal shocks, 
or to refreshed shocks, 
i.e. late time energy injections into the main afterglow shock by slow moving shells ejected from the central engine 
during the prompt phase.
The event fully satisfies the $E_{\rm{p}}$--$E_{\rm{iso}}$ Amati relation while is not consistent with 
the $E_{\rm p}$ vs. $E_{\rm jet}$ Ghirlanda relation.
}
{}

  \maketitle

\section{Introduction}
\label{intro}
\indent

The successful launch on 2004 November 20 of the \swift Gamma--ray Burst Explorer 
(\cite{gehrels}) has opened a new era in the study of Gamma Ray Bursts (GRBs). 
The autonomous and rapid slewing capabilities of \swift allow the prompt (1--2 minutes) 
observation of GRBs, discovered and localised by the wide-field gamma--ray (15--350 keV) 
Burst Alert Telescope (BAT, \cite{Barthelmy05a}), with the two co-aligned narrow-field instruments 
on-board the observatory: the X--Ray Telescope (XRT, \cite{Burrows05a}), operating in the 0.2--10 keV energy 
band, and the Ultraviolet/Optical Telescope (UVOT, \cite{roming}), sensitive in the 1700--6000 \AA ~band.
The \swift unique fast pointing capability is crucial in the X--ray energy band, where the reaction times of other 
satellites are limited to time scales of several hours. With {\it Swift}, thanks also to the XRT high sensitivity, it is 
possible for the first time to study in detail the evolution of the X--ray afterglows of GRBs during their 
very early phases.

Indeed, one of the main results of \swift is the identification of unexpected and complex features in the early 
X--ray afterglows. In particular, three distinct phases are observed in the majority of GRB light curves: 
an early ($t$$<$500 seconds from the trigger) steep decline, with a power-law index 
of $\sim$$\,3$, a second ($t$$<$10 ks) very shallow phase with a slope of $\sim$$\,0.5$, a third phase 
characterized by a more conventional decay slope of $\sim$$\,1$ 
(e.g. \cite{Tagliaferri05}; \cite{Cusumano06a}; \cite{Nousek06}; \cite{OBrien06}). In a few cases 
(GRB 050315, \cite{Vaughan06}; GRB 050318, \cite{Perri05}; GRB 050505, \cite{Hurkett06}; 
GRB 050525A, \cite{Blustin06}; GRB 060614, \cite{Mangano07}) a further steepening with a decay slope of $\sim$$\,2$, 
consistent with a jet break, is observed. Moreover, \swift had detected in about one half of the bursts 
strong flaring activity in the X--ray energy band superimposed on the afterglow decay 
(e.g. \cite{Burrows05b}; \cite{Romano06}; \cite{Falcone06}).

The understanding of the origin of these bright X--ray flares is intensively discussed 
in the literature. A mechanism proposed as responsible for the flaring activity is late internal shocks 
(\cite{Burrows05b}; \cite{Fan05}; \cite{Zhang06}; \cite{Liang06}). In this scenario the X--ray flares are produced 
by the same internal dissipation processes which cause the prompt emission, likely internal shocks within 
the expanding fireball occurring before it is decelerated in the external medium (e.g. \cite{Rees94}). 
This model requires that the GRB central engine is still active after the end of the prompt emission and 
various mechanisms providing such extended internal activity have been put forward 
(e.g. \cite{King05}; \cite{Perna06}).
An alternative scenario has been recently considered by Guetta et al.~(2007) who, based on a detailed analysis 
of the X--ray flaring activity observed in the afterglow of GRB 050713A, interpreted the 
X--ray flares as due to refreshed shocks, i.e. late time collisions with the 
main afterglow shock of slow-moving shells ejected from the central engine 
during the prompt phase (\cite{Rees98}; \cite{KumarPiran00}; \cite{Sari00}).

In this paper we present a detailed analysis of \swift and \xmm observations 
of GRB 050730, focusing on the intense and extended X--ray flaring activity that characterizes its afterglow. 
In Section \ref{obs} the observations and the data reduction 
are presented, in Section \ref{temporal} we describe the temporal analysis and 
Section \ref{spectral} is dedicated to the spectral analysis. The results are discussed 
in Section \ref{discussion} and finally in Section \ref{conclusions} we summarize our findings.
Throughout this paper errors are quoted at the 90\% confidence level for one  
parameter of interest ($\Delta\chi^{2}=2.71$) unless otherwise specified. 
We adopted the standard $\Lambda$CDM cosmological parameters of \mbox{$\Omega_\mathrm{m} = 0.27$}, 
\mbox{$\Omega_\Lambda = 0.73$} and \mbox{$H_0 = 70$~km~s$^{-1}$~Mpc$^{-1}$}. Times are 
referenced from the BAT trigger $T_0$ while temporal and spectral indices are written following 
the notation \mbox{$F(t,\nu) \propto t^{-\alpha} \nu^{-\beta}$}. Results on the optical spectrum 
of the afterglow of this GRB are reported by \cite{Starling05}, \cite{Chen05a} and \cite{Prochaska}. 
Multi-wavelength observations of the afterglow of GRB 050730 are presented by Pandey et al.~(2006).

   \begin{figure}[t]
   \centering
   \includegraphics[width=8.cm,angle=-90]{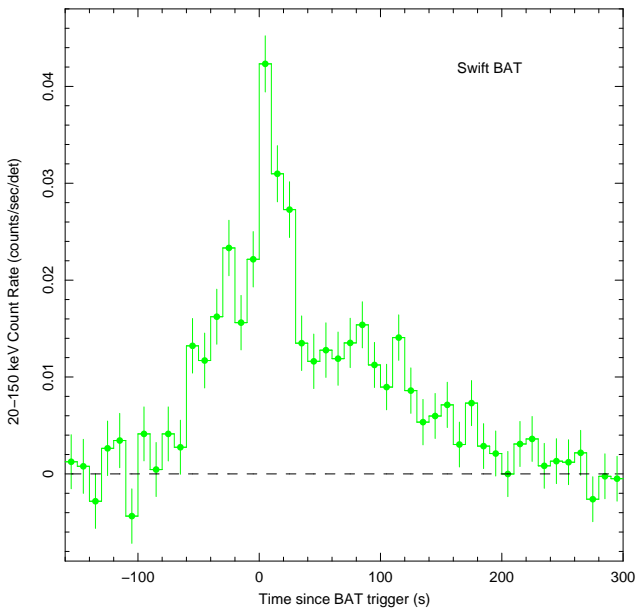}
      \caption{BAT 20--150 keV background subtracted light curve of the prompt emission 
               of GRB 050730. Data are binned to 10 seconds resolution and errors are 
               at the 1$\sigma$ level. The horizontal dashed line indicates the 0 level.
              }
         \label{bat_lc}
   \end{figure}

\section{Observations and data reduction}
\label{obs}
\subsection{Swift BAT}
\label{bat}
\indent

The BAT detected and located GRB 050730 at $T_0$=19:58:23 UT on 2005 July 30 (\cite{Holland05}). 
On the basis of the refined ground analysis (\cite{Mark05}), the BAT position is 
RA(J2000)=212\fdg063, Dec(J2000)=$-$3\fdg740, with a 90\% containment radius of 3\arcm. 
The BAT prompt emission light curve (Fig.~\ref{bat_lc}) is characterized by a duration of 
$\rm{T}_{90} = 155\pm20~\mathrm{s}$.  The emission starts $\sim$\,60 s before the trigger, peaks at 
$\sim$\,10 s after the trigger and declines out to $\sim$\,180 s after the trigger. 

The time averaged (over $\rm{T}_{90}$) spectrum in the 15--150 keV energy band is well described 
by a power-law model with energy index $\beta_{\rm BAT} = 0.5\pm0.1$ and $\chi_r^2=0.71$ (with 56 
degrees of freedom, d.o.f.). The total fluence in the 15--150 keV band is 
$(2.4\pm0.3)\times10^{-6}~\mathrm{erg}~\mathrm{cm}^{-2}$.
Assuming as redshift $z = 3.969$ (\cite{Chen05b}, see Sect.~\ref{ground}), the isotropic-equivalent radiated 
energy in the BAT bandpass (74.5--745.4 keV in the burst rest frame) is 
$E_{\rm iso}^{\rm BAT} = (8.0\pm 1.0) \times 10^{52}$~erg.

\subsection{Ground-based Observatories}
\indent
\label{ground}

Following the identification of the optical counterpart by UVOT (\cite{Holland05}), 
the field of GRB 050730 was observed by numerous ground-based telescopes.
The afterglow detections in the $R$ (\cite{Sota}), $r'$ (\cite{Gomboc}), $I$ and $J$ 
(\cite{Cobb}) bands were soon distributed via the GRB Circular Network (GCN). Observations 
in the optical band with the MIKE echelle spectrograph on Magellan II led to the GRB redshift 
measurement \mbox{$z=3.969$} (\cite{Chen05b}) based on the detection of a strong hydrogen 
Ly$\alpha$ absorption line and of several absorption lines from other elements. The redshift was later 
confirmed with observations by the ISIS spectrograph on the William Herschel Telescope (\cite{Rol}), 
the IMACS imaging spectrograph on the Magellan Observatory Baade Telescope (\cite{Holman05a}), 
FORS1 and UVES on the Very Large Telescope (\cite{Delia}).

The optical afterglow decay at later times was followed by several optical telescopes. In particular, 
measurements were made in the $R$ band up to $\sim$4 days after the trigger 
(\cite{Holman05a}, 2005b; \cite{Burenin}; \cite{Klotz}; \cite{Damerdji}; \cite{Delia}; 
\cite{Bhatt}; \cite{Kannappan}).

Finally, a radio afterglow with a flux density \mbox{$F_\mathrm{r}=145\pm28~\mu \mathrm{Jy}$} at 8.5 GHz 
was detected about 2 days after the trigger using the Very Large Array (\cite{Cameron}).

\subsection{Swift UVOT}
\indent
\label{uvot}

\swift UVOT began to observe the field of GRB 050730 at 20:00:22 UT, 119 seconds after 
the BAT trigger. The first 100 seconds exposure in the $V$ band led to the identification 
of the optical afterglow at \mbox{RA(J2000)=14$^{\rm{h}}$08$^{\rm{m}}17\fs09$}, 
\mbox{Dec(J2000)=-$03^{\circ} 46^{\prime} 18\farcs9$} (\cite{Holland05}). 

We refined the preliminary photometric analysis (\cite{Blustin05}) processing all UVOT data using the 
standard UVOT software package (\swift software v.~2.1 included in the HEAsoft package v.~6.0.2). 
The flux in all filters was 
estimated by integrating over a 3.5\arcs\ region. A background region for subtraction of the sky
contribution to the flux has been selected in a relatively empty part of the field of view.
The results are listed in Table~\ref{uvot_mag}.
No significant detection was found in the 
$U$ and $UV$ filters, which is consistent with the high redshift measured for this GRB.
All the magnitudes are corrected for Galactic extinction ($E(B-V)$ = 0.049, \cite{Schlegel}). 

\subsection{Swift XRT}
\label{xrt}
\indent

The XRT observations of the GRB 050730 field started at 20:00:28 UT, 125 seconds  
after the BAT trigger, with the instrument in Auto State. After a first exposure in 
Image Mode (see \cite{Hill} for a description of readout modes) during which no on-board 
centroid was determined, the instrument switched into Windowed Timing (WT) mode for the entire 
first \swift orbit from $T_0$+133~s to $T_0$+794~s. Starting from the second orbit ($T_0$+4001~s), 
the instrument was in Photon Counting (PC) mode for 26 consecutive 
orbits until 11:49:22 UT on 2005 August 1 ($T_0$+143.4~ks). The field of GRB 050730 was 
re-observed with the XRT from August 3 starting at 15:28:11 UT ($T_0$+329.4~ks) until 
August 5 13:57:02 UT ($T_0$+496.7~ks).

The XRT data were processed with the XRTDAS software (v.~1.7.1) included in the 
HEAsoft package (v.~6.0.4). Event files were calibrated and cleaned 
with standard filtering criteria with the \textsc{xrtpipeline} task using the latest calibration 
files available in the \swift CALDB distributed by HEASARC. Events in the energy range 
0.3--10 keV with grades 0--12 (PC mode) and 0--2 (WT mode) were used in the analysis 
(see \cite{Burrows05a} for a definition of XRT event grades).
After the screening, the total exposure time for the first XRT observation was 649 seconds 
(WT) and 58480 seconds (PC), while for the follow-up observation the PC exposure time was 
34669 seconds.

In the 0.3--10 keV PC image of the field a previously uncatalogued X--ray source was 
visible within the BAT error circle with coordinates \mbox{RA(J2000)=14$^{\rm{h}}$08$^{\rm{m}}17\fs2$}, 
\mbox{Dec(J2000)=$-03^{\circ} 46^{\prime} 19\arcs$}. 
This position, derived using data not affected by pile-up (orbits 5--26, see below), 
 has a 90\% uncertainty of 3.5\arcs\ using the latest XRT boresight 
correction (\cite{Moretti06}) and is consistent with the position of the optical counterpart 
(\cite{Holland05}; \cite{jacques05}).

For the WT mode data, events for temporal and spectral analysis were selected using a 40 pixel wide 
(1 pixel corresponds to $2.36\arcs$) rectangular region centered on the afterglow and aligned along the 
WT one dimensional stream in sky coordinates. Background events were extracted from a nearby source-free 
rectangular region of 50 pixel width. For PC mode data, the source count rate during 
orbits 2--4 was above $\sim$ 0.5 counts s$^{-1}$ and data were significantly affected by 
pile-up in the inner part of the Point Spread Function (PSF). After comparing the observed 
PSF profile with the analytical model derived by \cite{Moretti05}, we removed pile-up effects 
by excluding events within a 5 pixel radius circle centered on the afterglow position and used an outer 
radius of 30 pixels. From orbit 5 the afterglow brightness was below the pile-up limit and events 
were extracted using a 10 pixel radius circle, which encloses about 80\% of the PSF at 1.5 keV, to 
maximize the signal to noise ratio.
The background for PC mode was estimated from a nearby source-free circular region of 50 pixel radius. 
Source count rates for temporal analysis were corrected for the fraction of PSF falling outside 
the event extraction regions. Moreover, the loss of effective area due to the presence of 2 CCD 
hot columns within the extraction regions was properly taken into account.
The count rates were then converted into unabsorbed 0.3--10 keV fluxes using the conversion factor 
derived from the spectral analysis (see Sect.~\ref{spectral}). 

For the spectral analysis, ancillary response files were generated with the \textsc{xrtmkarf} task 
applying corrections for the PSF losses and CCD defects. The latest response 
matrices (v.~008) available in the \swift CALDB were used and source spectra were binned 
to ensure a minimum of 20 counts per bin in order to utilize the $\chi^{2}$ minimization 
fitting technique.

   \begin{figure*}[t]
   \centering
   \includegraphics[width=12cm,angle=-90]{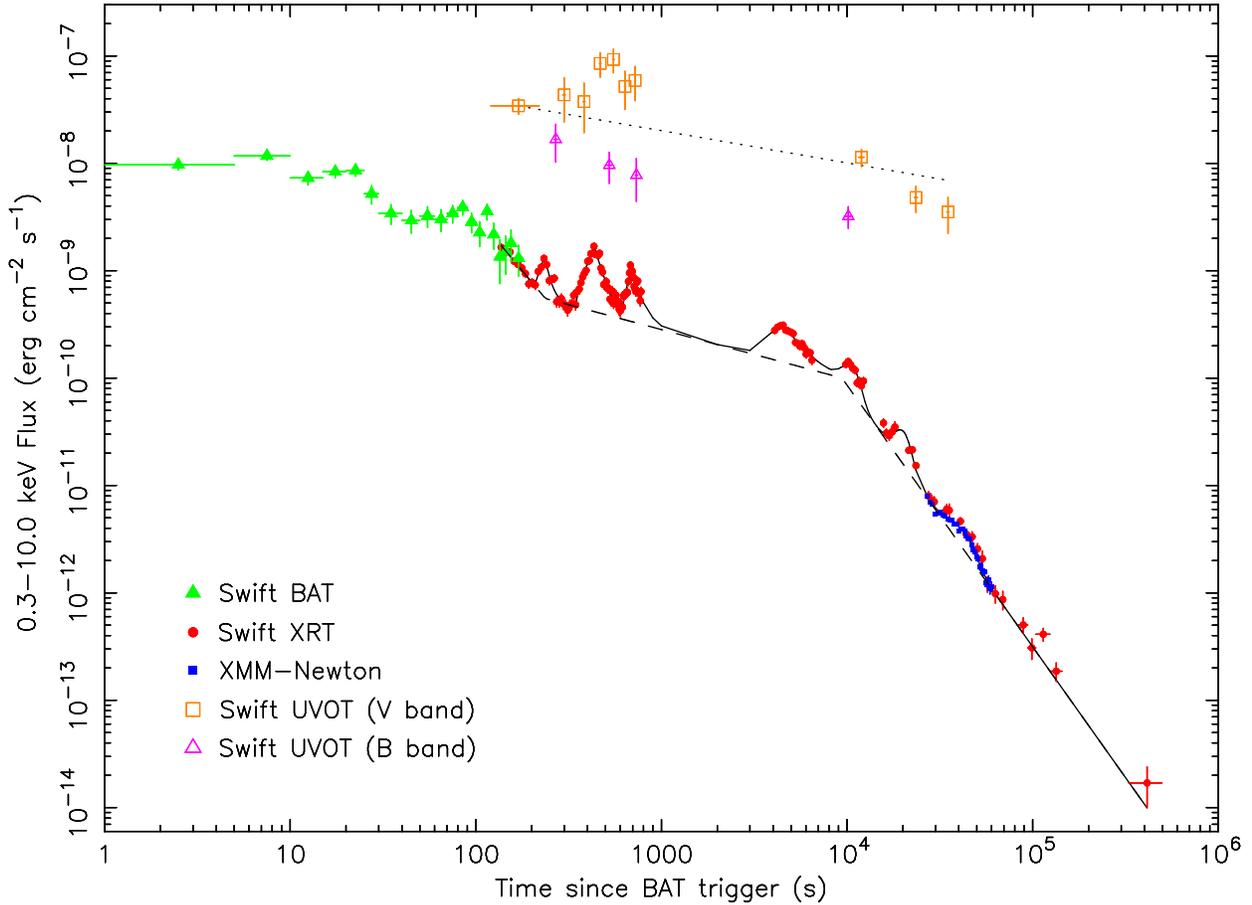}
      \caption{\swift XRT (filled circles) and \xmm (filled squares) 0.3--10 keV light curve 
               of the X--ray afterglow of GRB 050730. The BAT 20--150 keV prompt emission light curve, 
               extrapolated to the 0.3--10 keV band, is also shown as filled 
               triangles. UVOT optical data in the $V$ and $B$ bands, arbitrarily 
               scaled for comparison with the X--ray band, are indicated with open squares and 
               open triangles, respectively. The solid line is the best fit model to the XRT and \xmm 
               light curve. The dashed line represent the underlying double broken power-law 
               decay (see Sect.~\ref{early_xray}). 
               The dotted line is a power-law model with temporal decay index $\alpha_{\rm V}=0.3$ 
               normalized to fit data points in the $V$ band. Errors are at the 1$\sigma$ level.
              }
         \label{xrt_lc}
   \end{figure*}

\subsection{XMM-Newton}
\label{xmm}
\indent

\xmm\ follow-up observations of GRB 050730 started 26.4\,ks 
(for the two EPIC-MOS cameras) and 29.4\,ks (for the EPIC-PN) 
after the initial BAT Trigger. The \xmm\ ODF (Observation Data Files) 
data were processed with the \textsc{epproc} and \textsc{emproc} 
pipeline scripts, using the \xmm\ SAS analysis package (v.~6.5). 
A bright rapidly decaying source was detected 
near the aim-point of all three EPIC detectors and the afterglow was 
localized at \mbox{RA(J2000)=14$^{\rm{h}}$08$^{\rm{m}}17\fs3$}, 
\mbox{Dec(J2000)=-$03^{\circ} 46^{\prime} 18\farcs5$}.
The duration of the XMM-Newton follow-up observation was 33.7\,ks (MOS) and 
30.4\,ks (PN). After screening out times with high background flaring,  
the dead-time corrected net exposures were 25.0\,ks (MOS) and 17.9\,ks (PN).
All three EPIC cameras (PN and 2 MOS) were used in Full Window Mode, with 
PN and MOS2 using the ``Thin'' filter and MOS1 using the ``Medium'' 
optical blocking filter.

Source spectra and light curves for all 3 EPIC cameras were 
extracted from circular regions of 30\arcs\ radius centered on the
afterglow. Background data were taken from a 60\arcs\ radius
circle on the same chip as the afterglow, but free of any background 
X--ray sources. The data were further screened by including only good 
X--ray events (using the selection expression \textsc{flag=0} in evselect), 
by including events with single and double pixel events 
(\textsc{pattern<=4}) for the PN and by selecting single to quadruple pixel 
events (\textsc{pattern<=12}) for the MOS. Data below 300\,eV and above 
10\,keV were also removed. 

For the temporal analysis, we adopted the light curve from the MOS data, 
primarily because the PN data was heavily affected by background flares 
towards the end of the observation, while the MOS covered a wider duration 
at the beginning of the observation. The data from the two MOS detectors 
were combined and the count rate to 0.3--10 keV unabsorbed flux conversion 
factor was calculated from the best fit absorbed power-law spectrum (see 
Sect.~\ref{spectral}).  

For the spectral analysis, ancillary and redistribution response files for fitting
were generated with the SAS tasks \textsc{arfgen} and \textsc{rmfgen}, respectively. 
Moreover, source spectra were binned to a minimum of 25 counts per bin. 

\section{Temporal analysis}
\label{temporal}
\indent

The background subtracted 0.3--10 keV {\it Swift} XRT and \xmm light curves of the X--ray 
afterglow of GRB 050730 are shown in Fig.~\ref{xrt_lc}. The same figure also shows the 
BAT 20--150 keV prompt emission light curve, converted in the 0.3--10 keV energy band 
using the BAT spectral best fit model which is valid also in the XRT bandpass 
(see Sect.~\ref{spectral_xrt}). The UVOT optical light curves in the $V$ and $B$ bands, 
in arbitrary units, are also plotted (see Sect.~\ref{uvot}).

   \begin{figure}[t]
    \centering
    \includegraphics[width=8cm,angle=-90]{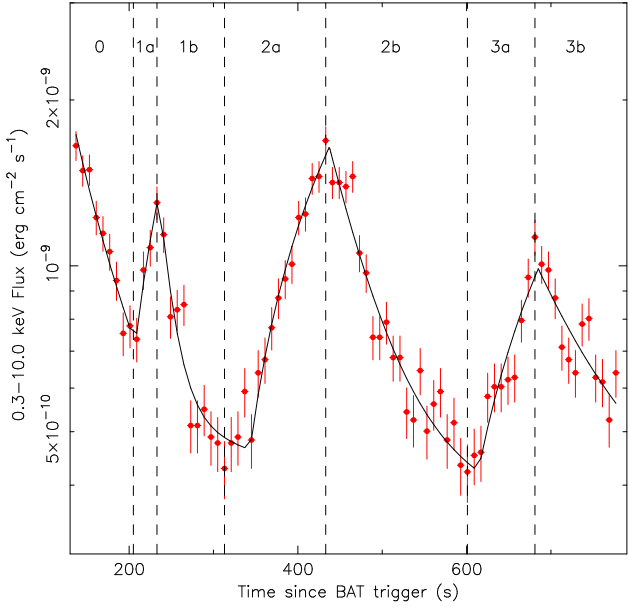}
      \caption{\swift XRT 0.3-10 keV light curve of the GRB 050730 X--ray afterglow during the 
               first orbit. The solid line is the best fit model to the data obtained considering 
               a linear rise exponential decay for the three flares (see Sect.~\ref{early_xray}).
               The dashed vertical lines delimit the seven time intervals 
               considered for the spectral analysis. Data are binned to 8 seconds resolution and errors are 
               at the 1$\sigma$ level.
              }
         \label{lc_flares}
   \end{figure}

\subsection{X--ray afterglow}
\label{early_xray}
\indent

The X--ray afterglow of GRB 050730 is characterized by a very complex structure. 
The first \swift orbit (from $T_0$+133~s to $T_0$+794~s), after an initial steep decay phase 
that joins well with the end of the BAT prompt emission, is dominated by three bright X--ray flares 
peaking at about 235, 435 and 685 seconds after the BAT trigger. Taking into account cosmological 
time dilation these times correspond to about 47, 88 and 138 seconds in the GRB rest frame. 
A flaring episode is also observed in the second orbit peaking at about $T_0$+4500 s. While the underlying 
decay of the afterglow during the first two orbits is shallow, starting from the third 
orbit ($T_0$+10~ks) the afterglow light curve shows a much steeper decline with superimposed 
flaring activity.

We first modeled the X--ray light curve of the afterglow with a double broken 
power-law model with slopes $\alpha_1$, $\alpha_2$, $\alpha_3$ and temporal breaks $t_1$, $t_2$, 
describing the underlying power-law decay of the afterglow, plus seven Gaussian functions 
modeling the flaring episodes. 
We found for the first power-law a decay index $\alpha_1 = 2.1\pm 0.3$ followed, 
after a first time break at $t_1 = 237\pm 20~\mathrm{s}$, by a shallower decay with index 
$\alpha_2 = 0.44_{-0.08}^{+0.14}$. A second temporal break is found at 
$t_2 = 10.1_{-2.2}^{+4.6}~\mathrm{ks}$ after which a steep decay with index 
$\alpha_3 = 2.40_{-0.07}^{+0.09}$ is observed. 
This model did not provide a good fit ($\chi_r^2=1.73$, 143 d.o.f.), 
mostly due to short time scale fluctuations and to deviations of the three first bright flares from 
a symmetric Gaussian shape.

   \begin{figure}[t]
   \centering
    \includegraphics[width=8cm,angle=-90]{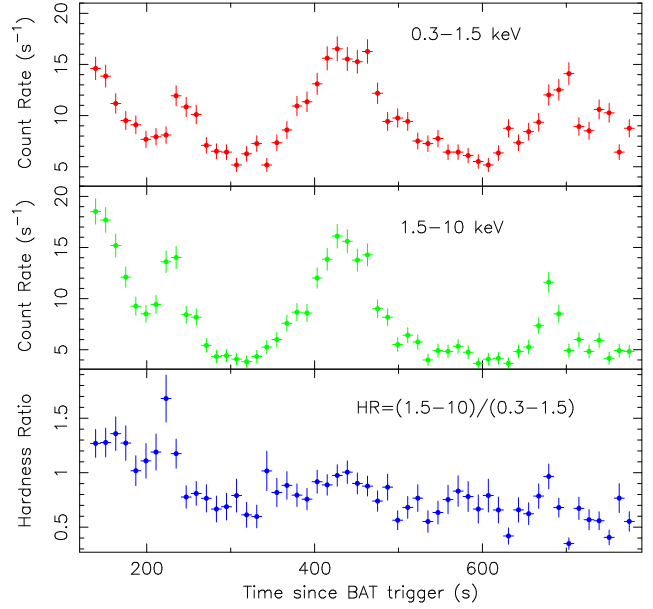}
     \caption{\swift XRT 0.3--1.5 keV (upper panel) and 1.5--10 keV (middle panel) light curve of the GRB 050730 
              X--ray afterglow during the first \swift orbit. In the lower panel the corresponding 
              hardness ratio is plotted. Data are binned to 12 seconds resolution and error bars 
              indicate statistical uncertainties at the 1$\sigma$ level.
             }
         \label{hardness}
   \end{figure}

We thus considered a different functional form for the three X--ray flares, namely a linear rise 
exponential decay: 
$F(t) \propto (t-t_0)/(t_{\rm p}-t_0)$ for 
times between the flare start time $t_0$ and peak time $t_{\rm p}$, 
\mbox{$F(t)\propto$ exp$[-(t-t_p)/t_{\rm c}]$} for 
$t>t_{\rm p}$ where $t_{\rm c}$ is the exponential decay time. The model improves significantly 
the fit with $\chi_r^2=1.43$ (140 d.o.f.) and F-test chance probability of $1.3\times 10^{-6}$. 
The best fit parameters of the overall underlying double broken power-law model were 
unchanged with respect to the previous fit listed above.
The linear rise exponential decay best fit parameters for the first three flares are given in Table 
\ref{flares_burs} while Table \ref{flares_gauss} reports the Gaussian best fits for the other four 
flares.
As Fig.~\ref{lc_flares} illustrates, the second brightest flare (referred to as flare 2 in the following) 
has a flux variation of amplitude $\sim 3.6$ and is characterized by a steep rise, 
lasting $\sim 90$ seconds, followed by a slower decay with duration $\sim 170$ seconds. An asymmetric shape, 
with a steep rise followed by a shallower decay, was also observed for the other two flares 
(flare 1 and 3) and for these episodes the flux variation was $\sim 1.7$ and $\sim 2.6$, 
respectively. 

The ratio between the duration ($\Delta t$) and the peak time ($t_{\rm p}$) of the X--ray flares 
was calculated using the linear rise 
exponential decay and Gaussian best fit parameters for the first three and the last four flares, 
respectively. For both analytical models the duration of a flare was defined as the time 
interval during which the flare intensity was above 5\% of its peak value. The results are 
given in Tables \ref{flares_burs} and \ref{flares_gauss}.

The rise and decay portions of the first three flares were also fit with single power-laws. For 
the estimation of power-law indices, times were expressed from the onset of the flares 
($t_0$) for the rise, from the peak time ($t_{\rm p}$) for the decay, 
and the contribution of the underlying afterglow decay was taken into account. 
The power-law temporal indices for the rising ($\alpha_{\rm r}$) and 
decaying ($\alpha_{\rm d}$) portions of the three flares are reported in Table \ref{flares_pow}.

The temporal behavior of the afterglow during the first three X--ray flares was also studied in 
different energy bands. In Fig.~~\ref{hardness} we show the 0.3--1.5 keV (upper panel) and 
1.5--10 keV (middle panel) light curves of the early X--ray afterglow together with the 
corresponding hardness ratio (lower panel). From the figure it is apparent that i) in the harder band 
the profile of the flares is sharper and ii) a peak time shift, with flares in the hard band peaking 
at earlier times, is observed. 
We note that these temporal properties as a function of energy are the same observed for the prompt 
emission pulses of GRBs.

\begin{figure}[t]
   \centering
   \includegraphics[width=8cm,angle=-90]{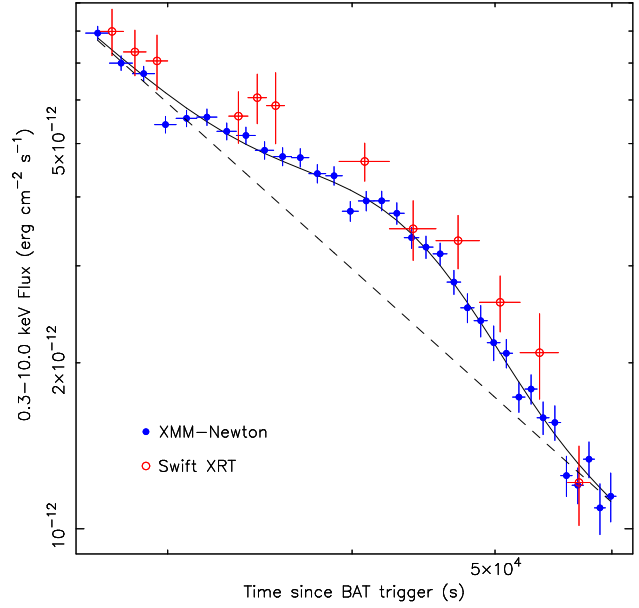}
      \caption{\xmm (MOS1+MOS2, filled circles) and \swift XRT (open circles) 0.3--10 keV light curve 
               of the late ($\sim$ $T_0$+12 hours) X--ray afterglow of GRB 050730. 
               The solid line is the best fit model to the \xmm 
               light curve. The dashed line represent the underlying power-law decay 
               (see Sect.~\ref{early_xray}). Data points errors are at the 1$\sigma$ level.
              }
         \label{xmm_lc}
   \end{figure}

As already mentioned, late flaring activity of the X--ray afterglow 
was observed at $\sim$ 4.5, 10.4, 18.7 and 41.2 ks 
after the trigger. Due to the \swift orbital gaps the temporal coverage of flares 4, 5 and 6 
is poor and these episodes could not be well studied, but the last flare was entirely covered by the 
\xmm follow-up observation allowing a detailed temporal analysis.
In Fig.~\ref{xmm_lc} the \xmm (MOS1+MOS2) light curve is shown together with the XRT curve 
covering the same time interval. 
A very good agreement between the two curves is found and the $\sim 5\%$ higher normalization 
of the XRT data points is of the same order of the uncertainties in the absolute flux 
calibration of the instrument (\cite{Campana}).

We first fit the \xmm light curve with a single power-law decay and obtained an extremely 
poor fit to the light curve, with a fit statistic of 
$\chi^{2}_r=8.81$ (32 d.o.f.) and a decay index of $\alpha=2.10\pm0.04$. 
The light curve was then parameterized with a long duration flare super-imposed on a steep power-law decay. 
We obtained an underlying decay index $\alpha=2.45\pm0.15$, while the 
flare could be adequately modeled with a Gaussian 
function; the fit statistic was acceptable ($\chi^{2}_r=1.22$, 29 d.o.f.). 
The long duration flare peaked at 41\,ks (or 8.2\,ks in the GRB rest frame), 
with a $\sigma$ of 6.8\,ks (1.4\,ks rest frame). The total fluence of the flare 
when parameterized this way was $8.3\times10^{50}$\,erg (1.5-50\,keV 
band in the GRB rest frame), which represents 20\% of the integrated afterglow 
emission over the \xmm observation.
The rise and decay phases of flare 7 were also fit with single power-laws expressing times from 
the onset and from the peak time of the flare, respectively. The measured temporal 
indices $\alpha_{\rm r}$ and $\alpha_{\rm d}$ are also reported in Table \ref{flares_pow}.

\subsection{Optical band}
\label{optical_curve}
\indent

The UVOT optical light curve of the afterglow of GRB 050730 in the $V$ and $B$ bands is also 
characterized by a complex behavior, as is illustrated in Fig.~\ref{xrt_lc}.

The early ($T_0$+100~s - $T_0$+800~s) UVOT $V$ light curve reveals flaring activity. 
A re-brightening is in fact observed at $\sim T_0$+500~s, almost simultaneously 
with the brightest X--ray flare observed with XRT. Due to the relatively poor sampling 
of the UVOT light curve, a detailed temporal analysis is not possible, however we note that the amplitude 
of the flux variation in the $V$ band ($\sim 3$) is of the same order of the one measured in the X--ray 
energy band. The decay of the afterglow in the $V$ band, up to about $T_0$+12~ks and excluding the 
re-brightening episode, is shallow ($\alpha_{\rm V} \sim 0.3$) and in agreement with the one measured 
in the X--ray band ($\alpha_{\rm X}=0.44_{-0.08}^{+0.14}$). 
Hints of a steepening of the $V$ light curve at $\sim T_0$+12~ks are also found, but the limited 
temporal coverage and the large statistical uncertainties characterizing the UVOT data points do not allow 
us to constraint the late optical afterglow decay.

The UVOT light curve in the $B$ band is characterized by a flat power-law decay ($\alpha_{\rm B} \sim 0.3$) without 
re-brightening events. However, it should be noted that i) the curve is very sparsely sampled, with only 
three data points during the first \swift orbit, and ii) in the $B$ band there is a strong flux reduction due 
to the Lyman break and thus large statistical uncertainties affect the data points. 

Accurate optical observations of the afterglow of GRB 050730 in the $R$ and $I$ bands are presented by 
Pandey et al.~(2006). The authors report an early time flux decay with indices 
$\alpha_{\rm R} = 0.54\pm 0.05$ and $\alpha_{\rm I} = 0.66\pm 0.11$ followed, after a temporal break at about 
9 ks after the trigger, by a strong steepening with decay indices 
$\alpha_{\rm R} = 1.75\pm 0.05$ and $\alpha_{\rm I} = 1.66\pm 0.07$.

\section{Spectral analysis}
\label{spectral}
\indent
For the spectral analysis of \swift XRT and \xmm data we used the XSPEC package (v.~11.3.2, \cite{Arnaud96}) 
included in the HEAsoft package (v.~6.0.4).

\subsection{XRT}
\label{spectral_xrt}
\indent

As a first step, the 0.3--10 keV XRT average spectrum during the first \swift orbit 
(WT mode, from $T_0$+133~s to $T_0$+794~s) was fit adopting a single power-law model 
with the neutral hydrogen-equivalent absorption column density fixed at the Galactic value in the direction 
of the GRB ($N_{\rm H}^{\rm G} = 3.0\times10^{20}$\,cm$^{-2}$, \cite{Dickey90}). 
The fit obtained was poor ($\chi^2_{\rm r}=1.28$, 311 d.o.f.). From the inspection of residuals a strong deficit 
of counts at low energies, likely due to the presence of an absorbing column density in excess to the Galactic one, 
was found. Indeed, the addition of a column density $N_{\rm H}^z$ using Solar metallicity 
redshifted to the rest frame of the GRB host ($z=3.969$) free to vary in the spectral fit 
({\sc zwabs} model in {\sc XSPEC}) resulted in a more acceptable fit with a column density 
of \mbox {$N_{\rm H}^z=(1.28^{+0.26}_{-0.25})\times 10^{22}$\,cm$^{-2}$} (see Table~\ref{powerlaw}). 
This value is consistent 
with the neutral hydrogen column densities derived from the optical spectra reported by \cite{Starling05} and 
\cite{Chen05a}. 

From Fig.~\ref{hardness} it is apparent that strong spectral evolution takes place during the intense flaring 
activity observed in the first \swift orbit. We thus split the WT observation 
in seven time intervals to study the spectra during the rise and the decay portions of each flare 
(see Fig.~\ref{lc_flares}). 
The time-resolved spectral best fits (see Table~\ref{powerlaw}) clearly show evidence for spectral variation during 
the flares and an overall softening of the 
spectra with time associated with a decrease of the rest frame column density. 
The time-resolved spectral analysis was also performed adopting a broken power-law model to investigate the possible 
presence of spectral breaks (e.g. \cite{Falcone06}; \cite{Guetta07}). Also in this case an 
additional absorption column density $N_{\rm H}^z$ at the rest frame of the GRB host was considered. 
For all segments we did not find evidence for spectral breaks within the XRT energy band.

The XRT late 0.3--10 keV spectrum (PC mode, from $T_0$+4.0~ks to $T_0$+143.4~ks) was also fit using a single 
power-law model with a fixed Galactic absorption column density and an additional absorbing column at the 
burst rest frame. The observation was divided in two time intervals (4.0--18.1 ks and 21.3--143.4 ks 
from the trigger). The results are listed in Table~\ref{powerlaw} where we see that spectral softening 
between the two time intervals was found.

\subsection{XMM-Newton}
\indent

The PN and MOS 0.3--10 keV spectra were fit jointly allowing the cross normalization between the detectors, 
which is consistent within $<5\%$, free to vary . The two MOS spectra and responses were combined to 
maximize the signal to noise, after first checking that they were consistent with each other. 

As in the case of the XRT spectrum, the PN and MOS spectra were first fit with a single power-law model with the 
neutral hydrogen-equivalent absorption column fixed at the known Galactic value. 
The fit obtained was not acceptable ($\chi^2_{\rm r}=1.40$, 489 d.o.f.), while the energy index was 
$\beta=0.76\pm0.02$. 
The addition of a neutral absorption column in the GRB host galaxy frame at $z=3.969$ 
resulted in more acceptable fit ($\chi^{2}_{\rm r}=1.14$, 489 d.o.f.) with 
an excess absorption column above the Galactic column of $(6.8\pm 1.0)\times10^{21}$\,cm$^{-2}$, 
while the continuum energy index was now $\beta=0.87\pm0.02$.

The \xmm\ afterglow spectra were also sliced into three segments of 
approximately 10\,ks in duration, in order 
to search for any spectral evolution within the \xmm\ observation. 
A small change in the continuum parameters was found, the spectrum 
evolved from hard to soft; the energy index 
changed from $\beta=0.87\pm0.03$ in the first 10\,ks, 
to $\beta=0.99\pm0.05$ during the final segment. No evidence was found for a change in the 
column density, which was subsequently fixed at 
$N_{\rm H}=6.8\times10^{21}$\,cm$^{-2}$ in all the segments. The spectral 
best fit parameters are shown in Table~\ref{powerlaw}.

We also searched for any evidence of emission lines, either in the mean 
spectrum, or in the three segments. No statistically significant lines were 
found, at the level of $>99\%$ confidence. As the redshift of the 
burst is known, then we can set an upper-limit to the equivalent width 
of any emission lines. Over the range of 0.4--8 keV and using the 
mean spectrum, we found a $<100$\,eV 
upper limit to any emission lines. More specifically we can set a 
limit on the iron K$\alpha$ line (e.g. the H-like line at 6.97\,keV 
rest frame, 1.40\,keV observed frame) of $<30$\,eV.  

\section{Discussion}
\label{discussion}
\indent

\subsection{Early X--ray light curve}
\label{early}
\indent

This GRB, with its rather high redshift, $z=3.969$, gives us the possibility to 
investigate the X--ray and optical light curves soon after the trigger and thus to study in detail the soft tail of the 
prompt emission and the very beginning of the afterglow phase. Indeed, due to cosmological time dilation 
the XRT and UVOT observations started, in the rest frame of the burst, only 27 seconds after the trigger. 
Other examples of such early observations of high redshift bursts are GRB 050904 ($z=6.29$, \cite{Cusumano06b}) and 
GRB 060206 ($z=4.0$, \cite{Monfardini06}).

The XRT light curve shows at the very beginning (133--205 s from the BAT trigger) a rapidly decaying 
emission that joins quite nicely with the BAT flux when converted to the XRT bandpass, a feature that has been 
observed in various \swift bursts (\cite{Tagliaferri05}, \cite{Barthelmy05b}; \cite{Nousek06}; \cite{OBrien06}).
For GRB 050730, the steep temporal decay index ($\alpha_1 = 2.1\pm 0.3$, see Sect.~\ref{early_xray}) 
characterizing the early X--ray light curve suggests that the observed emission is likely associated with the 
tail of the prompt emission rather than to the beginning of the shocked inter-stellar medium afterglow 
phase. Indeed, in the internal shock model scenario for the prompt $\gamma$--ray emission 
(e.g. \cite{Rees94}), the cessation of the emissivity is characterized by a rapid decay 
due to the delayed arrival of the high angular latitude prompt emission of the shocked surface 
(high-latitude or curvature emission, \cite{KumarPanaitescu00}; \cite{Dermer04}). 
The high-latitude emission predicts a relationship between the 
temporal decay index $\alpha_{\rm d}$ and the spectral index $\beta_{\rm d}$ during the decay given by 
the equation $\alpha_{\rm d}=2+\beta_{\rm d}$ (\cite{KumarPanaitescu00}) where the decay slope is measured using as 
zero time ($t_0$) the beginning of the prompt emission 
(\mbox{$F(t,\nu) \propto (t-t_0)^{-\alpha_{\rm d}} \nu^{-\beta_{\rm d}}$}).
We have tested this prediction by fitting the early XRT light curve (segment ``0'' in Fig.~\ref{lc_flares}) 
with the above constraint using the spectral index measured during the decay 
($\beta_{\rm d} = 0.42\pm 0.08$, see Table \ref{powerlaw}) and leaving the zero time as a free parameter 
(see \cite{Liang06} for a description of the method). We found $t_0=-31^{+24}_{-31}$ s, i.e. the zero time 
is located at the rising segment of the BAT prompt emission light curve (see Fig.~\ref{bat_lc}), as expected 
in the framework of the high-latitude emission. This result indicates that the steep decay observed in the early 
XRT light curve is most likely the soft tail of the BAT emission.

This hypothesis is also supported by the BAT and early (segment ``0'' in Fig.~\ref{lc_flares}) 
XRT spectra. We found very similar spectral slopes 
($\beta_{\rm BAT} = 0.5\pm0.1$ and $\beta_{\rm XRT} = 0.42\pm0.08$, respectively), likely indicating that 
the early X--ray and $\gamma$--ray emissions are produced by the same emission mechanism.

\subsection{Flaring activity}
\label{flaring}
\indent

The exceptionally extended flaring activity of the X--ray afterglow of GRB 050730 allows us to study this 
phenomenon over more than two orders of magnitude in time. 
The XRT early steep decay is followed by three bright X--ray flares peaking at 236, 437 and 685 
seconds after the BAT trigger (see Fig.~\ref{lc_flares}). These flares were under the sensitivity of the 
BAT instrument and thus were not detected in the hard X--ray energy band.
These flares show, as other strong flares observed with \swift (e.g. \cite{Romano06}; \cite{Falcone06}; 
\cite{Pagani06}; \cite{Godet06}; \cite{Chincarini06}), a clear spectral evolution with the hardness ratio that mimics 
the variation of the light curve. A phase lag, with the harder ($E>1.5$ keV) light curve peaking 
at earlier times with respect to the softer energy band is also found (see Fig.~\ref{hardness}).

So far, most of the X--ray flares observed in \swift X--ray light curves have been interpreted as due to late 
internal shocks (e.g. \cite{Burrows05b}; \cite{Romano06}; \cite{Falcone06}; \cite{Godet06}). In this scenario 
the GRB central engine is active far beyond the end of the prompt $\gamma$--ray emission phase requiring new 
mechanisms capable of powering new relativistic outflows at late-time (e.g. \cite{King05}; \cite{Perna06}).
A diagnostic to check if the re-brightenings are due to late internal shocks has been recently proposed by 
Liang et al.~(2006). In the internal-origin scenario for X--ray flares the decay emission of 
the flaring episodes should be dominated by high-latitude emission with the decay temporal 
index related to the decay spectral index as $\alpha_{\rm d}=\beta_{\rm d}+2$ (\cite{KumarPanaitescu00}).
We have checked this hypothesis for the first three bright X--ray flares observed in the GRB 050730 afterglow. The 
spectral indices measured during the decays were used (segments ``1b'', ``2b'' and ``3b'', 
see Fig.~\ref{lc_flares} and Table \ref{powerlaw}) and we found best fit zero times 
$t_{0,1}=$193$^{+8}_{-11}$ s, $t_{0,2}=$341$^{+29}_{-42}$ s and $t_{0,3}=$592$^{+35}_{-93}$ s, 
respectively. 
We can see that for all three episodes the zero time values are located at the beginning 
of the corresponding flare and thus the observed decay slopes are consistent with being due to 
high-latitude emission as predicted by the late internal shock scenario.

Another possibility is that the X--ray flares are produced by refreshed shocks (\cite{Rees98}; 
\cite{KumarPiran00}; \cite{Sari00}). In the standard internal-external fireball model (e.g. \cite{Rees94}; \cite{Sari97}) 
re-brightening episodes in the afterglow light curve are explained as slow shells ejected from the central 
engine during the prompt phase that catch up with the main 
afterglow shock after it has decelerated in the external inter-stellar medium. Indeed, Guetta et al.~(2007) 
have recently interpreted the X--ray flares of the afterglow of GRB 050713A as due to refreshed shocks. 
Here we applied the diagnostic proposed for GRB 050713A to the case of GRB 050730. First, to estimate the rise 
and decay slopes of the early flares we operationally selected as zero time the beginning and the 
peak of the flare, respectively. This assumption implicitly means that the flare is completely 
independent from the main event generating the prompt emission and the forward shock light curve. We found 
temporal indices in the 0.8--1.8 range (see Table \ref{flares_pow}). Much steeper slopes ($\sim$ 3--6) are obtained 
if instead times are referenced from the BAT trigger, as usually done for most of the \swift X--ray flares 
(e.g. \cite{Burrows05b}; \cite{Romano06}; \cite{Falcone06}).
Second,  we checked if the temporal and the spectral indices during the decay phase are related as predicted 
by the standard afterglow model (\cite{Sari98}).
We restricted the analysis to the episodes which have enough statistics to allow a detailed temporal 
and spectral study, i.e. the first three flares and the last one. For the decay phases we expect the 
following relations to hold (\cite{Sari98}; \cite{Dai01}):
$F_\nu \propto
(\nu_m/\nu_c)^{-1/2}(\nu/\nu_m)^{-p/2} F_{\nu,max} \propto
t^{-3p/4+1/2}$ (for $p>2$) and $F_\nu \propto
(\nu_m/\nu_c)^{(p-1)/2}(\nu/\nu_c)^{-p/2} F_{\nu,max} \propto
t^{-(3p+10)/16}$ (for $1<p<2$).
From the best fit spectral energy indices during the decay of the four flares (see Table \ref{powerlaw}) 
we derived the predicted temporal slopes $\alpha_{\mathrm d}^1 = 0.93\pm 0.05$, $\alpha_{\mathrm d}^2 = 0.89\pm 0.03$, 
$\alpha_{\mathrm d}^3 = 1.02\pm 0.15$, and $\alpha_{\mathrm d}^7 = 1.40\pm 0.05$, 
which are consistent with the measured values of the decay indices listed in Table \ref{flares_pow}. 
We note that for the last flare, peaking at 41.2 ks after the trigger, we used the afterglow model relations for the slow 
cooling case ($F_\nu \propto (\nu/\nu_m)^{-(p-1)/2} F_{\nu,max} \propto
t^{3(1-p)/4}$) which very likely applies at these late times (\cite{Sari98}).

Important information on the mechanism producing the flares can also be obtained from the comparison between the 
observed variability timescale and the time at which the flare is observed (\cite{Ioka05}).
We calculated, for all the 7 re-brightenings of the GRB 050730 afterglow, the ratio between the duration of the 
flare ($\Delta t$) 
and the time when the flare peaks ($t_{\rm p}$) (see Tables \ref{flares_burs} and \ref{flares_gauss}).
We found $\Delta t/t_{\rm p} \sim 0.3$ for the first flare and $\Delta t/t_{\rm p} \sim 0.6-0.7$ for the others, 
in agreement with the $\Delta t/t_{\rm p}>0.25$ limit discussed by Ioka et al.~(2005) for the refreshed shock 
scenario. 
Remarkably, the $\Delta t/t_{\rm p}$ ratio is nearly constant for all flares, with the exception of the first 
one which occurred during the bright and steep tail of the prompt emission (see Fig.~\ref{lc_flares}) 
and most likely we are observing only the tip of the flare. A duration of the flare proportional to $t_{\rm p}$, as 
observed in the afterglow of GRB 050730, is explained both in the refreshed shock scenario (\cite{KumarPiran00}) 
and in the late internal shock model (\cite{Perna06}).
We note that the last four flares have $\Delta t/t_{\rm p}$ values in the range 0.7--0.9, while for 
flares 2 and 3 we observe slightly lower ratios (0.5--0.6), possibly indicating a moderate temporal evolution of 
$\Delta t/t_{\rm p}$. However, due to the relatively large error bars, the measured values are 
also consistent with being constant and $\sim 0.6-0.7$.

Flaring activity is also observed in the optical afterglow of GRB 050730. Indeed, the UVOT $V$ early light curve 
(see Fig.~\ref{xrt_lc}) shows a re-brightening at $\sim T_0$+500~s, almost simultaneously 
with the brightest X--ray flare observed with the XRT peaking at $T_0$+437~s. Moreover, the optical 
re-brightening amplitude ($\sim 3$) is of the same order of the flux variation observed in the X--ray energy band. 
Strong indications of correlated variability between the X--ray and optical energy bands are also present in the 
NIR/Optical light curves reported by Pandey et al.~(2006): a significant re-brightening at about $T_0$+4~ks 
($R$ filter), close to the X--ray flare peaking at $T_0$+4.5~ks (see Fig.~\ref{xrt_lc}), is observed. 
Moreover, a bump in 
the $J$, $I$, $R$ and $V$ light curves at about $T_0$+10~ks is observed (\cite{Pandey06}), again simultaneous with 
the X--ray flare peaking at $T_0$+10.4~ks. Although the X--ray flaring activity is not uniformly 
covered by optical observations, we find several indications of simultaneous re-brightening events in the X--ray and 
optical bands, in agreement with the refreshed shock model (e.g. \cite{Granot03}).

\subsection{Evidence of a jet break}
\label{break}
\indent

The X--ray afterglow light curve shows a clear temporal break around 10~ks after the trigger 
(see Sect.~\ref{early_xray}). At about the same time a strong steepening of the $I$ and $R$ afterglow light curves, 
although with a shallower post-break slope, is also observed (see \cite{Pandey06} 
for a detailed analysis on the discrepancy between the X--ray and optical slopes).
Due to the achromatic nature of the temporal break it is very likely that a jet break is occurring 
when the bulk Lorentz factor $\gamma$ of the collimated relativistic outflow becomes lower than the inverse of 
the jet opening angle $\theta_{\rm jet}$ (e.g., \cite{Rhoads97}, \cite{Sari99}), as also reported by 
Pandey et al.~(2006).
In this framework, the jet opening angle  can be determined  through the equation 
$\theta_{\rm jet} = 0.161 [t_{\rm b}/(1+z)]^{3/8}(n \eta/E_{\rm iso})^
{1/8}$ (e.g., \cite{Bloom03}) where $\theta_{\rm jet}$ is in radians, 
the jet temporal break $t_{\rm b}$ in days, the total isotropic-equivalent energy 
$E_{\rm iso}$ in units 10$^{52}$ erg, the density $n$ of the circumburst medium  in cm$^{-3}$ 
and $\eta$ is the efficiency of conversion of the outflow kinetic energy in electromagnetic radiation. 

An accurate estimation of the bolometric isotropic-equivalent energy radiated by GRBs requires the 
knowledge of their intrinsic spectrum over a broad energy band (\cite{Bloom01}; \cite{Amati02}). 
For GRB 050730, the spectrum is well fit by a single power-law with energy index $\beta_{\rm BAT} = 0.5\pm0.1$ 
up to the high energy limit of the BAT sensitivity bandpass (150 keV) indicating that i) we are observing 
the low energy tail of the Band model (\cite{Band93}) generally used to describe 
GRB spectra and ii) the $\nu F(\nu)$ spectrum peak energy $E_{\rm p}$ 
is above $\sim 750$ keV (i.e. $(1+z)\times 150$ keV) in the burst rest frame. A lower limit to the 
bolometric isotropic-equivalent radiated energy $E_{\rm iso}$ is given by the observed radiated energy in 
the BAT bandpass ($E_{\rm iso}^{\rm BAT} = (8.0\pm 1.0) \times 10^{52}$~erg, see Sect.~\ref{bat}). 
An upper limit to $E_{\rm iso}$ is obtained in the most conservative case where the peak energy 
$E_{\rm p}$ is equal to $10^4$ keV (\cite{Amati02}). By integrating the best fit BAT 
power-law spectrum in the whole 1--$10^4$ keV rest frame energy band we thus derived an upper limit of 
$4.5\times 10^{53}$~erg to $E_{\rm iso}$.
Taking the central value of the interval derived above 
we obtained for GRB 050730 $E_{\rm iso} = (2.6\pm 1.9) \times 10^{53}$~erg.

With $z=3.969$, $t_{\rm b} = 10.1_{-2.2}^{+4.6}~\mathrm{ks}$ (see Sect.~\ref{early_xray}) and 
the $E_{\rm iso}$ range above derived, we find $\theta_{\rm jet} = 1.6^{+0.6}_{-0.2}$ deg, for 
$\eta = 0.2$ (\cite{Frail01}) and assuming the value of circumburst 
density $n = 10$~cm$^{-3}$ discussed by Bloom et al.~(2003). With this value of the 
jet opening angle, the inferred collimation-corrected bolometric radiated energy is 
$E_{\rm jet} = (1.0^{+2.3}_{-0.8}) \times 10^{50}$~erg. 

Taking into account the $E_{\rm iso}$ and $E_{\rm jet}$ values derived above and the lower limit 
to the peak energy ($E_{\rm p}>$ 750 keV rest frame), we 
find that GRB 050730 is consistent with the $E_{\rm p}$ vs. $E_{\rm iso}$ relation found by Amati et al.~(2002) 
and recently updated in Amati (2006). We also find that GRB 050730 is inconsistent, even taking into 
account the 3$\sigma$ scatter around the best fit correlation, with the $E_{\rm p}$ vs. $E_{\rm jet}$ 
relation found by Ghirlanda et al.~(2004) and with its updated version presented by Nava et al.~(2006). 
In order to make this GRB consistent with the $E_{\rm p}$-$E_{\rm jet}$ relation a much higher 
circumburst density ($n \sim 10^5$~cm$^{-3}$) would be required.
GRB 050730 is inconsistent with the model-independent $E_{\rm iso}$-$E_{\rm p}$-$t_{\rm b}$ correlation 
found by Liang \& Zhang (2005).

\section{Summary and conclusions}
\label{conclusions}
\indent

We have presented a detailed temporal and spectral analysis of the afterglow of GRB 050730 observed 
with \swift and {\it XMM-Newton}. 
The most striking feature of this GRB is the intense and exceptionally extended, over more than two order 
of magnitude in time, X--ray flaring activity.

Superimposed to the afterglow decay we observed seven distinct re-brightening events peaking 
at 236 s, 437 s, 685 s, 4.5 ks, 10.4 ks, 18.7 ks and 41.2 ks after the BAT trigger. 
The underlying decline of the afterglow was well described with a double broken power-law model with breaks at 
$t_1 = 237\pm 20~\mathrm{s}$ and $t_2 = 10.1_{-2.2}^{+4.6}~\mathrm{ks}$. The temporal decay slopes before, between 
and after these breaks were $\alpha_1 = 2.1\pm 0.3$, $\alpha_2 = 0.44_{-0.08}^{+0.14}$ and 
$\alpha_3 = 2.40_{-0.07}^{+0.09}$, respectively. 

Strong spectral evolution during the flares was present together with an overall softening of the underlying 
afterglow with the energy index varying from $\beta=0.42\pm 0.08$ during the early (133--205 s) steep decay to 
$\beta=0.99\pm 0.05$ at much later (50-60 ks) times. An absorbing column density 
$N^{z}_\mathrm{H}=(1.28^{+0.26}_{-0.25})\times10^{22}$\,cm$^{-2}$ in the host galaxy is observed during the early 
(133--781 s) \swift observations while a lower column density $N^{z}_\mathrm{H}=(0.68\pm0.10)\times10^{22}$\,cm$^{-2}$ 
is measured during the late (29.4--50.8 ks) \xmm follow-up observation, likely indicating photo-ionization of the 
surrounding medium. Evidence of flaring activity in the early UVOT optical afterglow, simultaneous with that 
observed in the X--ray band, was found. 

From the temporal analysis of the first three bright X--ray flares we found that the rise and decay power-law slopes 
are in the range 0.8--1.8 if the beginning and the peak of the flares are used as zero time, respectively.
We also found that, with the exception of the first flare, for all episodes the ratio between the duration 
of the flare ($\Delta t$) and the time when the flare peaks ($t_{\rm p}$) is nearly constant and is 
$\Delta t/t_{\rm p}\sim 0.6-0.7$.

We showed that the observed properties of the first three flares are consistent with being due to both 
high-latitude emission, as expected for flares produced by late internal shocks, or to late time energy 
injection into the main afterglow shock by slow moving shells (refreshed shocks).
An analysis of a larger sample of bursts would help in understanding what are the physical mechanisms 
responsible for the X-ray flaring activity.

We interpreted the X--ray temporal break at around 10~ks as a jet break and derived a cone 
angle of $\sim$\,2 deg and a radiated energy 
\mbox{$E_{\rm jet}=(0.2-3.3)\times 10^{50}$~erg} against an isotropic-equivalent energy 
\mbox{$E_{\rm iso}=(0.7-4.5)\times 10^{53}$~erg}. 
GRB 050730 satisfies the $E_{\rm p}$ vs. $E_{\rm iso}$ Amati relation 
while is inconsistent with the $E_{\rm p}$ vs. $E_{\rm jet}$ Ghirlanda relation.

\begin{acknowledgements}

We are grateful to the referee for his/her useful comments and suggestions. We also thank 
C. Guidorzi for a very careful reading of the paper and 
F. Tamburelli and B. Saija for their work on the XRT data reduction software.
This work is supported in Italy from ASI on contract number I/R/039/04 and through funding 
of the ASI Science Data Center, at Penn State by NASA contract NAS5-00136 and at the 
University of Leicester by the Particle Physics and Astronomy Research Council on grant 
numbers PPA/G/S/00524 and PPA/Z/S/2003/00507.
\end{acknowledgements}

\begin{table*}
\caption{UVOT detections of GRB 050730 in the $V$ and $B$ filters. Column (1) gives the image 
         mid time in seconds since the BAT trigger, column (2) the net exposure time, 
         column (3) the filter used, and column (4) the afterglow magnitude with 1$\sigma$ error.
        }
\label{uvot_mag}     
\centering                  
\begin{tabular}{c c c c }    
\hline\hline                
T(mid) &  Exposure & Filter & Magnitude \\  
(s) &  (s) &  &  \\  
\hline                   
170     & 99    & $V$ & $17.4\pm 0.2$   \\
270     & 10    & $B$ & $18.4\pm 0.4$   \\
299     & 9.7   & $V$ & $17.2\pm 0.5$   \\      
383     & 9.7   & $V$ & $17.3\pm 0.5$   \\      
468     & 9.7   & $V$ & $16.4\pm 0.3$   \\
524     & 30    & $B$ & $19.0\pm 0.4$   \\
552     & 9.7   & $V$ & $16.3\pm 0.3$   \\      
637     & 9.7   & $V$ & $17.0\pm 0.4$   \\      
721     & 9.7   & $V$ & $16.8\pm 0.4$   \\
734     & 20    & $B$ & $19.2\pm 0.5$   \\
10164   &  900  & $B$ & $20.2\pm 0.3$   \\
11947   & 837   & $V$ & $18.6\pm 0.2$   \\
23519   & 835   & $V$ & $19.5\pm 0.3$   \\
34990   & 843   & $V$ & $19.9\pm 0.4$   \\
\hline
\end{tabular}
\end{table*}

\begin{table*}
\caption{Temporal best fit parameters of the first three bright X--ray flares of GRB 050730 using 
         a linear rise exponential decay model. The corresponding $\Delta t/t_{\rm p}$ values for 
         the three flares are also indicated (see Sect.~\ref{early_xray}).
        }
\label{flares_burs}     
\centering                  
\begin{tabular}{l c c c c c}    
\hline\hline                
Flare   & $t_0$         &  $t_{\rm p}$      & $t_{\rm c}$  &  $K\times 10^{-10}$    & $\Delta t/t_{\rm p}$ \\  
          &     (s)       &  (s)              &  (s)         & (\ergcms) &              \\
\hline                    
1 & $207^{+5}_{-5}$   & $236^{+3}_{-4}~~$ & $17^{+6}_{-5}$   & $8\pm 1$       & $0.3\pm 0.1$   \\
2 & $344^{+5}_{-5}$   & $437^{+5}_{-4}~~$ & $54^{+6}_{-6}$   & $12.2\pm 0.7$  & $0.58\pm 0.04$ \\
3 & $614^{+~9}_{-10}$ & $685^{+7}_{-6}~~$ & $87^{+37}_{-21}$ & $6.1\pm 0.7 $   & $0.5\pm 0.1$   \\
\hline
\end{tabular}
\end{table*}

\begin{table*}
\caption{Temporal best-fit parameters of the X--ray flares 4, 5, 6 and 7 of GRB 050730 using Gaussian functions.
         An asterisk indicates a frozen parameter. The corresponding $\Delta t/t_{\rm p}$ values 
         are also indicated (see Sect.~\ref{early_xray}). Best-fit parameters for flare 7 were derived from \xmm 
         data only.
        }
\label{flares_gauss}     
\centering                  
\begin{tabular}{l c c c c}    
\hline\hline                
Flare  & $t_{\rm p}$    &  $\sigma$    & $K\times 10^{-11}$        & $\Delta t/t_{\rm p}$ \\
          &     (s)   &  (s)         & (\ergcms) & \\ 
\hline                    
4 &  $4484^{+124}_{-235}$ &  $641^{+259}_{-149}$   & $11\pm 2$         & $0.7\pm 0.2$   \\
5 & $10391^{+330}_{-315}$ & $1500^*~~~~~~~$        & $5.5\pm 1.0$    & $0.71\pm 0.02$ \\
6 & $18714^{+952}_{-1156}$& $3638^{+820}_{-789}$   & $1.5\pm 0.4$    & $0.9\pm 0.2$   \\ 
7 & $41244^{+691}_{-744}$ & $6170^{+819}_{-718}~~$ & $0.12\pm 0.02$ & $0.7\pm 0.1$   \\ 
\hline
\end{tabular}
\end{table*}

\begin{table*}
\caption{ Best fit temporal indices of the rising ($\alpha_{\rm r}$) and decaying ($\alpha_{\rm d}$) portions 
         of the X--ray flares 1, 2, 3 and 7 using a single power-law model.
        }
\label{flares_pow}     
\centering                  
\begin{tabular}{l c c}    
\hline\hline                
Flare &$\alpha_{\rm r}$&$\alpha_{\rm d}$\\
\hline                    
1 & $-1.56\pm 0.69$ & $1.25\pm 0.32$ \\
2 & $-1.84\pm 0.30$ & $1.31\pm 0.44$ \\
3 & $-1.46\pm 0.62$ & $0.75\pm 0.35$ \\
7 & $-0.89\pm 0.34$ & $1.61\pm 0.44$ \\
\hline
\end{tabular}
\end{table*}

\begin{table*}[h]
\caption{Results of single power-law spectral fits to the 0.3--10 keV spectrum of the afterglow of GRB 050730. 
         A local ($z=0$) absorption column fixed at the known Galactic value 
         of $N_{\rm H}^{\rm G} = 3.0\times10^{20}$\,cm$^{-2}$ (\cite{Dickey90}) was used in the fits. 
         An asterisk indicates a frozen parameter.
        }
\label{powerlaw}     
\centering                  
\begin{tabular}{l c l c l}    
\hline\hline                
segment   & time interval &  $N^{z}_\mathrm{H} \times 10^{22}$ & $\beta$ & $\chi_r^2$ (d.o.f.) \\  
          &   (s)       &          (cm$^{-2})$         &         &                     \\
\hline                    
WT (all)  &  133--781     &  $1.28^{+0.26}_{-0.25}$ & $0.70^{+0.03}_{-0.03}$ & 1.01 (310) \\
          &               &                         &                        &            \\
WT (0)    &  133--205     &  $1.8^{+0.9}_{-0.8}$    & $0.42^{+0.08}_{-0.08}$ & 0.86 (76)  \\
WT (1a)   &  205--233     &  $1.6^{+2.5}_{-1.6}$                 & $0.29^{+0.16}_{-0.16}$ & 0.86 (24)  \\
WT (1b)   &  233--313     &  $3.1^{+1.3}_{-1.1}$    & $0.82^{+0.12}_{-0.12}$ & 1.30 (51)  \\
WT (2a)   &  313--433     &  $2.1^{+0.8}_{-0.7}$    & $0.71^{+0.08}_{-0.08}$ & 1.10 (93)  \\
WT (2b)   &  433--601     &  $0.9^{+0.5}_{-0.5}$    & $0.70^{+0.07}_{-0.07}$ & 1.11 (111) \\
WT (3a)   &  601--681     &  $0.7^{+0.8}_{-0.7}$                 & $0.77^{+0.12}_{-0.12}$ & 0.87 (46)  \\
WT (3b)   &  681--781     &  $1.0^{+0.6}_{-0.6}$    & $1.01^{+0.10}_{-0.10}$ & 0.81 (65)  \\
          &               &                         &                        &            \\
PC (1)    & 4001--18149   &  $1.1^{+0.4}_{-0.4}$    & $0.61^{+0.04}_{-0.04}$ & 0.95 (224) \\
PC (2)    & 21288--143438 &  $1.0^{+0.7}_{-0.6}$    & $0.81^{+0.08}_{-0.08}$ & 1.16 (86)  \\
          &               &                         &                        &            \\
XMM (all) & 29436--59811  &  $0.68^{+0.10}_{-0.10}$ & $0.87^{+0.02}_{-0.02}$ & 1.14 (489) \\
          &               &                         &                        &            \\
XMM (1)   & 29436--40000  &  $0.68^*$               & $0.87^{+0.03}_{-0.03}$ & 0.88 (345) \\
XMM (2)   & 40000--50000  &  $0.68^*$               & $0.93^{+0.03}_{-0.03}$ & 0.98 (243)    \\
XMM (3)   & 50000--59811  &  $0.68^*$               & $0.99^{+0.05}_{-0.05}$ & 0.77 (135)    \\

\hline
\end{tabular}
\end{table*}

\end{document}